\documentclass[twocolumn, aps, prapplied, floatfix, nobibnotes]{revtex4-2}

\usepackage{amsmath}
\usepackage{amssymb}
\usepackage{graphicx}
\usepackage{hyperref}
\usepackage{xcolor}
\usepackage{siunitx}
\usepackage{enumitem}
\hypersetup{
    colorlinks=true,
    linkcolor={blue!80!black},
    citecolor={blue!80!black},
    urlcolor={blue!80!black}
}

\DeclareSIUnit\rad{rad}

\begin{document}
\title{Quantitative Rules for Parallel Calibration and Drift-Resilient Maintenance of Large-Scale Superconducting Quantum Processors}
\author{Zhihao Wu}
\thanks{These authors contributed equally to this work.}
\author{Kechengqi Zhu}
\thanks{These authors contributed equally to this work.}
\author{Shichuan Xue}
\email{shichuanxue@quanta.org.cn}
\author{Xinfang Zhang}
\author{Zhaohua Yang}
\author{Zhida Zhang}
\author{Shun Hu}
\author{Cong Li}
\author{Dingdong Liu}
\author{Kangding Zhao}
\author{Peng Luo}
\author{Lingling Lao}
\author{Xiang Fu}
\author{Mingtang Deng}
\email{mtdeng@nudt.edu.cn}
\author{Junjie Wu}
\author{Kai Lu}
\email{kailu@nudt.edu.cn}

\affiliation{College of Computer Science and Technology, National University of Defense Technology, Changsha $410073$, China}
\begin{abstract}
As superconducting quantum processors scale toward hundreds of qubits, manual calibration becomes increasingly costly and difficult to sustain. Here we address two system-level factors governing automated calibration, namely, crosstalk induced by parallel operations and temporal drift of control parameters. On a 66-qubit superconducting processor, cross-entropy benchmarking shows that single-qubit calibration tolerates full-chip concurrency with minimal fidelity loss, whereas CZ calibration requires a minimum coupler-graph distance of four to avoid elevated control errors. Continuous monitoring of representative qubit and coupler parameters further yields quantitative refresh rules that assign refresh priority by error-budget consumption. These results turn two conventionally heuristic choices, the concurrency density of calibration tasks and the refresh priority of each control parameter, into measurable operating rules. Implemented within an automated calibration framework, the rules support a six-qubit Greenberger-Horne-Zeilinger state with a zero-noise-extrapolated fidelity of $89.64\%$, distance-dependent logical-error suppression in a single-cycle repetition-code benchmark, and a reduction of the mean single-qubit error from $0.023$ to $0.011$ on a 337-qubit processor, providing an experimentally grounded reference for the automated calibration and maintenance of current and future large-scale superconducting quantum systems.
\end{abstract}

\date{\today}
\maketitle

\section{Introduction}

Superconducting circuits provide a scalable and programmable architecture for quantum information processing~\cite{blais2021circuit,kjaergaard2020superconducting}. Experimental progress has advanced from quantum computational advantage based on random circuit sampling~\cite{arute2019quantum,wu2021strong} to the suppression of logical errors with increasing surface-code distance~\cite{google2023suppressing,Google2024QEC,he2025experimental} and to the demonstration of logical-qubit operations and lattice surgery on superconducting hardware~\cite{besedin2026lattice, wang2026superconducting, lin2026surface}. As processors scale up, however, the number and interdependence of control parameters, calibration experiments, and validation tasks expand rapidly~\cite{klimov2024optimizing,mohseni2024build}. Establishing an operational processor from initial parameter estimates requires a coordinated sequence of readout, qubit, and coupler calibrations. 

\begin{figure*}[ht]
\includegraphics[width=1 \linewidth]{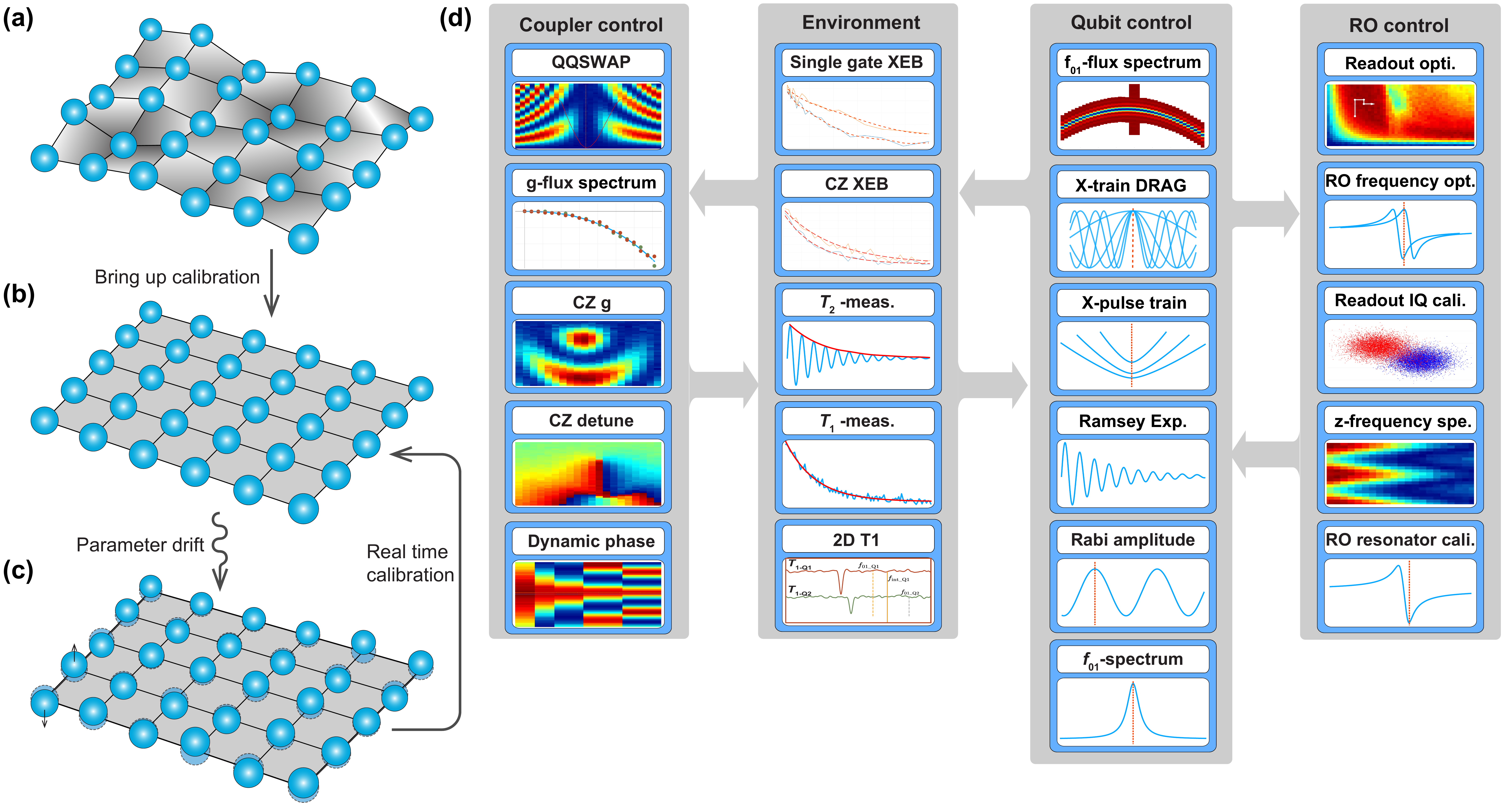}
\caption{\label{fig:qopilot_workflow}
A typical full-cycle calibration and maintenance workflow of a large-scale superconducting quantum processor, illustrated on the 66-qubit device used in this work. (a) Roughly estimated parameters during the initial cold start. (b) Calibration to achieve a high-fidelity state. (c) Random parameter drift caused by environmental fluctuations. (d) Directed acyclic graph (DAG) of the calibration experiments, containing four core modules, namely, readout control, qubit control, coupler control, and environment characterization. The detailed workflow is described in Appendix~\ref{app:workflow}.
}
\end{figure*}

As shown in Fig.~\ref{fig:qopilot_workflow}, a typical calibration and maintenance workflow starts from roughly estimated parameters at the initial cold start [Fig.~\ref{fig:qopilot_workflow}(a)], from which the processor is brought to a high-fidelity operating state [Fig.~\ref{fig:qopilot_workflow}(b)] through the dependency-ordered calibration experiments of Fig.~\ref{fig:qopilot_workflow}(d). During subsequent operation, variations in device and control conditions can displace the calibrated operating point [Fig.~\ref{fig:qopilot_workflow}(c)], so a maintenance workflow must update parameters before accumulated calibration offsets materially degrade circuit performance.

Automated calibration has advanced rapidly. Scheduling on directed acyclic graphs (DAGs) selects which calibration experiments to repeat~\cite{DAG}, and local optimization accelerates the assignment of interdependent controls~\cite{localop}; deep reinforcement learning has produced gates that remain robust against drift~\cite{deeplearning}; and large-language-model agents now execute multistep calibration or characterization workflows~\cite{cao2025kagents, li2026large, xu2026vibe}. FPGA-driven feedback tracks fluctuating parameters in real time~\cite{berritta2026, marciniak2026}, reaching millisecond timescales on superconducting qubits; automatic recalibration has sustained unattended 20-qubit operation for more than 100 days~\cite{mansfield2025first}; and reinforcement learning has managed over 1000 control parameters and further reduced distance-5 surface-code logical error rates by about $20\%$~\cite{sivak2025}. Complementary frameworks and surveys reflect this shift toward algorithmically driven control~\cite{alexeev2025, xu2021qubic, wittler2021integrated, xu2022automatic, pasquale2023towards}. DAG-based calibration has been applied on processors of order $10^{2}$ qubits, and maintaining optimal performance deep below the error-correction threshold is expected to require model-free updates alongside targeted metrology~\cite{sivak2025}.

Despite this progress, two fundamental and interrelated challenges continue to constrain the efficacy of large-scale continuous calibration, namely, multi-body crosstalk induced by highly parallel operations and temporal drift of physical control parameters (qubit, coupler, and readout settings). Although both issues are widely recognized as critical bottlenecks in the literature~\cite{kosen2024crosstalk,li2026systematic,mohseni2024build}, systematic quantitative investigation of their impact on large-scale calibration fidelity and throughput remains limited.

First, as massively parallel automated calibration becomes operationally indispensable for processors with a large number of qubits, the choice of an appropriate concurrency density becomes acute. Parallel execution inevitably introduces crosstalk, which manifests as frequency collisions among simultaneously driven qubits and residual $ZZ$-type or flux-mediated interactions between proximate gates.
Such crosstalk degrades calibration accuracy and, if uncontrolled, can render parameter determination unreliable. Yet the quantitative relationship among parallelism level, crosstalk severity, and ultimate calibration fidelity has not been established experimentally, leaving practitioners without principled guidance for balancing throughput against fidelity.

Second, parameter drift is an unavoidable consequence of cryogenic temperature fluctuations, two-level system dynamics, and evolution of the electromagnetic environment. The most commonly monitored quantities include qubit frequency $\omega_{01}$, energy relaxation time $T_1$, and decoherence time $T_2$. Beyond these, the temporal evolution of many control-relevant parameters, such as readout resonator frequencies, coupler biases, and crosstalk compensation matrices, has been insufficiently characterized. Without a quantitative classification of drift magnitudes and timescales across the full parameter set, there is no systematic basis for deciding whether a global recalibration, a targeted parameter refresh, or a predictive model-based update is most appropriate. This absence of a stratified drift-management strategy represents a critical open challenge for sustained fault-tolerant operation.

In this work, we systematically investigate crosstalk induced by parallel calibration operations and temporal drift of control parameters. We demonstrate that single-qubit gates tolerate full-chip concurrency with only a small fidelity loss, whereas two-qubit CZ gates require a minimum coupler-graph distance of four ($d_{\mathrm{C}}=4$) to suppress flux crosstalk. From continuous monitoring, we further classify key control parameters into three drift tiers of high, medium, and low priority and derive a stratified refresh strategy that allocates calibration resources by measured drift severity. Implemented in an automated calibration framework, these rules enable the preparation of a six-qubit Greenberger-Horne-Zeilinger (GHZ) state with a peak zero-noise-extrapolated state-tomography fidelity of $89.64\%$, distance-dependent suppression of logical errors in a single-cycle qubit-flip repetition-code benchmark on a 66-qubit processor, and a factor-of-two reduction of the average single-qubit Pauli error on a 337-qubit processor. More broadly, these results show that the two key operational choices of automated calibration, namely the concurrency density of calibration tasks and the refresh priority of each control parameter, can be determined quantitatively from measurements on the processor itself rather than set by heuristic convention.

\section{Parallelized Calibration}
\label{sec:parallel_calibration}
This work was carried out on a 66-qubit superconducting processor with a two-dimensional tunable-coupler architecture in flip-chip packaging, integrating 66 transmon qubits, 109 tunable couplers, and 11 readout feedlines~\cite{koch2007transmon, yan2018tunable, wallraff2004strong}. The qubit working-point frequencies span 3.9 to 4.8 GHz and the readout resonators lie between 6.2 and 6.6 GHz; further device details are given in Ref.~\onlinecite{wu2025telegraph}. All calibration tasks follow the workflow of Fig.~\ref{fig:qopilot_workflow}, organized as a directed acyclic graph of readout, qubit, and coupler calibration experiments and executed by the QoPilot automated calibration framework, with the full cold-start and maintenance sequence detailed in Appendix~\ref{app:workflow}. The parallelism and drift measurements presented below are carried out within this workflow.

Serial execution is a bottleneck at this scale. A full single-qubit calibration suite (frequency, derivative-removal-by-adiabatic-gate (DRAG) pulse shaping, readout) can require \SI{30}{\minute} per qubit, or $\sim\SI{50}{\hour}$ of wall-clock time for a 100-qubit processor, which makes frequent recalibration impractical for quantum error correction (QEC), where physical gate errors must remain below threshold for extended periods. The module dependency graph of the workflow (Appendix~\ref{app:workflow}) encodes the conditional independence among calibration tasks and thereby enables safe parallelization. In this section, we quantitatively characterize the resulting fidelity-speed tradeoff and identify the optimal parallel densities for single- and two-qubit gate calibrations.

\begin{figure}[!htp]
\centering
\includegraphics[width=0.95\linewidth]{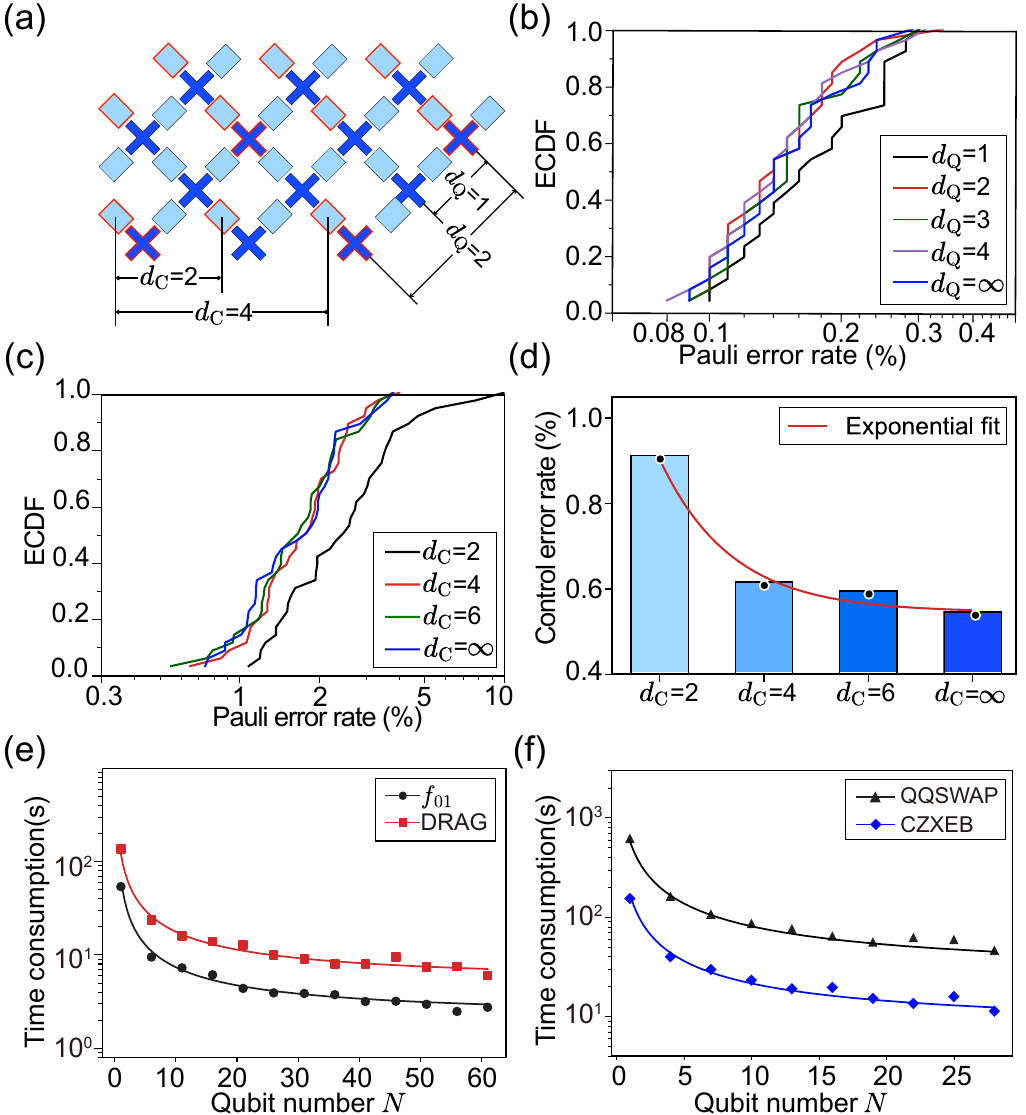}

\caption{
Topological definition of parallel density and calibration performance evaluation.
(a)~Schematic illustration of parallel densities $d_{\mathrm{Q}}$ and $d_{\mathrm{C}}$ defined as the minimum graph distance between simultaneously active qubits or couplers, respectively, on the processor topology.
(b)~Empirical cumulative distribution function (ECDF) of cross-entropy benchmarking (XEB) Pauli error rates for single-qubit gates at five values of $d_{\mathrm{Q}}$ (1,2,3,4,$\infty$).
(c)~ECDF of cross-entropy benchmarking (XEB) Pauli error rates for two-qubit CZ gates at four values of $d_{\mathrm{C}}$ (2,4,6,$\infty$).
(d)~Mean control error rate averaged over all calibrated CZ gates at each of the four coupler parallel densities in~(c), with the exponential distance dependence overlaid.
(e)~Execution time of single-qubit calibration experiments ($f_{01}$ spectroscopy and DRAG coefficient optimization) versus parallel grouping size $N$.
(f)~Execution time of two-qubit calibration experiments (QQSWAP and CZXEB) versus $N$.
}
\label{fig:parallel_calibration}
\end{figure}


\subsection{Single-qubit gate parallelism}

To quantify the susceptibility of single-qubit gates to concurrent control, we performed simultaneous single-qubit cross-entropy benchmarking (XEB) on qubit groups selected by the minimum lattice distance $d_{\mathrm{Q}}$, after an initial rough turn-off of the tunable couplers based on the qubit-qubit SWAP (QQSWAP) operation~\cite{yan2018tunable}. The resulting XEB Pauli error distributions under five parallel densities are compared in Fig.~\ref{fig:parallel_calibration}(b). From isolated operation ($d_{\mathrm{Q}}=\infty$) to full-chip concurrent driving ($d_{\mathrm{Q}}=1$), the empirical cumulative distribution functions (ECDFs) remain closely overlapping and the errors of all measured qubits remain below $0.4\%$. Small residual differences between the distributions remain discernible, but the additional error introduced by concurrent driving is much smaller than the variation of the measured single-qubit error distribution and is therefore negligible.

This weak dependence on $d_{\mathrm{Q}}$ can be understood qualitatively from the response of a spectator qubit to microwave leakage. Crosstalk between neighboring $XY$ control channels acts as an unintended transverse drive whose effect depends on the leakage amplitude, qubit frequency detuning, and relative pulse phase. A near-resonant component can perturb the effective rotation amplitude and axis, whereas an off-resonant component suppresses direct population transfer and mainly produces a weak transient AC Stark shift~\cite{zhao2022quantum, wang2022control}. An analytical model of this leakage mechanism is given in Appendix~\ref{app:microwave_crosstalk}. In the random sequences used for XEB, these pulse-level perturbations are sampled over different control axes and phases, limiting systematic coherent accumulation in the extracted average Pauli error. Dedicated microwave routing further attenuates direct coupling between adjacent control channels, while the QQSWAP-based turn-off biases the tunable couplers near their idle points and suppresses residual exchange and static $ZZ$ interactions. Since the present figure does not include a channel-resolved microwave crosstalk measurement, this mechanism is used only as a qualitative interpretation of the observed XEB robustness. We therefore adopt $d_{\mathrm{Q}}=1$ for single-qubit calibration, since the associated fidelity loss is small compared with the qubit-to-qubit variation.

\subsection{Two-qubit gate parallelism}

The fidelities of CZ gates calibrated under four coupler parallel densities are compared in Fig.~\ref{fig:parallel_calibration}(c), where the calibration performance is reported through the ECDFs of the XEB Pauli error rates. The curves for $d_{\mathrm{C}}=4$, $6$, and $\infty$ closely overlap, indicating that the sparse parallel configurations preserve performance comparable to isolated calibration. In contrast, the distribution at $d_{\mathrm{C}}=2$ shifts toward higher error rates, demonstrating that the denser configuration degrades the calibrated CZ gates.

To isolate the contribution associated with control, we calculated the control error of each calibrated CZ gate as the difference between its XEB error rate and its speckle purity benchmarking error rate~\cite{arute2019quantum}, and then averaged this quantity over all CZ gates at each parallel density. The resulting mean control errors are compared in Fig.~\ref{fig:parallel_calibration}(d), together with an exponential fit. The mean increases to approximately $0.91\%$ at $d_{\mathrm{C}}=2$, whereas the values for $d_{\mathrm{C}}=4$, $6$, and $\infty$ are $0.61\%$, $0.59\%$, and $0.54\%$, respectively, close to the isolated-calibration baseline. Fitting the distance dependence with an exponential crosstalk model,
\begin{equation}
\epsilon_{\rm ctrl}(d_{\mathrm{C}})=\epsilon_{\rm base}+A e^{-d_{\mathrm{C}}/\lambda},
\end{equation}
gives $\epsilon_{\rm base}=0.54\%$, $A=1.58\%$, and a decay length $\lambda=1.37$ graph edges. 

The CZ gate is implemented with single-polarity rectangular flux pulses, and parasitic mutual inductance or shared return paths can transfer part of a neighboring pulse into the target Z-channel~\cite{rol2019fast, kosen2024crosstalk}. The resulting bias offset and waveform distortion shift the instantaneous qubit or coupler frequencies, perturb the effective detuning and coupling strength, and consequently modify both the accumulated conditional phase and the leakage suppression condition. The effective Hamiltonian analysis in Appendix~\ref{app:flux_crosstalk} quantifies this sensitivity. For the operating parameters considered here, a coupler frequency perturbation of $\delta\omega_c/2\pi\approx\SI{1}{\mega\hertz}$ produces a conditional phase error of $\delta\Theta\approx\SI{0.14}{\rad}$ and an average gate fidelity penalty of approximately $\SI{0.3}{\percent}$, providing a quantitative explanation for the additional average control error observed at $d_{\mathrm{C}}=2$. At $d_{\mathrm{C}}\geq 4$, the larger spatial separation attenuates the mutual inductive coupling, and the XEB distribution and mean control error recover to the isolated calibration level. The combined results therefore identify $d_{\mathrm{C}}=4$ as the densest tested configuration that preserves CZ calibration performance comparable to isolated operation.

\subsection{Execution time scaling and classical control bottlenecks}

The fidelity analysis above determines how densely calibration tasks may be packed without unacceptable loss; whether this density translates into a practical advantage depends on the time it actually saves. We therefore measured the duration of four representative calibration experiments, namely single-qubit $f_{01}$ spectroscopy, DRAG coefficient optimization, QQSWAP spectroscopy, and CZ-gate XEB (CZXEB), as a function of the parallel grouping size $N$ [Figs.~\ref{fig:parallel_calibration}(e) and \ref{fig:parallel_calibration}(f)]. The per-task time follows an Amdahl-like form~\cite{amdahl1967validity},
\begin{equation}
U(N)=b+\frac{T_{\rm seq}-b}{N},
\end{equation}
where $T_{\rm seq}$ is the single-task (sequential) duration and $b$ is the marginal overhead per additional concurrently calibrated task. The fits give $b=1.86$~s/task ($f_{01}$), $4.66$~s/task (DRAG), $29.94$~s/task (QQSWAP), and $7.89$~s/task (CZXEB), corresponding to asymptotic speedups of approximately $29\times$ for single-qubit calibrations and $20\times$ for two-qubit calibrations. The saturation at high concurrency reflects the serial overhead $b$ from pulse preparation, instrument communication, data acquisition, parameter estimation, and feedback updates.

\subsection{Parallel-density rules and summary}

Synthesizing the fidelity and timing analyses, we adopt $d_{\mathrm{Q}}=1$ for single-qubit calibration because full concurrent driving barely changes the measured XEB Pauli-error distribution. For two-qubit calibration, $d_{\mathrm{C}}=4$ is the densest tested configuration for which both the XEB distribution and the mean control error remain comparable to isolated calibration. At these parallel densities, the timing measurements demonstrate substantial but sublinear experiment-level speedup with saturation at high concurrency. The value of $d_{\mathrm{C}}=4$ therefore provides an operating rule for the present processor layout, pulse implementation, and experimental control resources.

\begin{figure*}[ht]
\centering
\includegraphics[width=\textwidth]{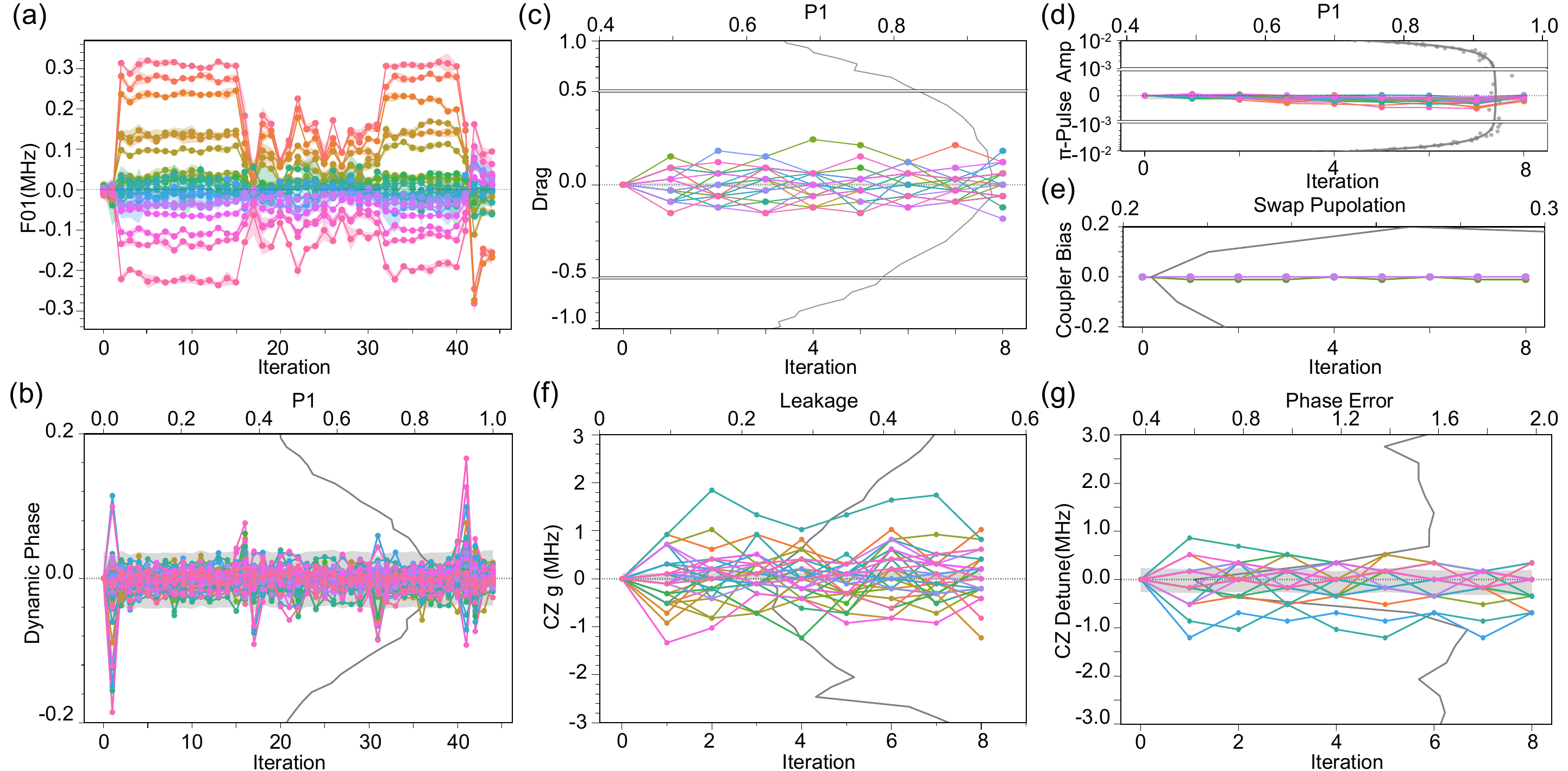}
\caption{
Monitoring results of key control parameters over $44$ and $48$ hours. (a) Qubit frequency $f_{01}$. (b) Dynamic phase. (c) DRAG coefficient. (d) $\pi$-pulse amplitude. (e) Coupler idle bias. (f) CZ effective coupling strength $g$. (g) CZ detuning. Panels (a) and (b) span 44 sampling iterations; panels (c) to (g) contain nine representative calibration iterations. Colored trajectories represent absolute parameter offsets from initial values of different qubits or couplers. The shaded band in (a) indicates the $1\sigma$ Ramsey repeatability. The upper horizontal axes and gray curves show the corresponding single-run calibration responses.
}
\label{fig:parameter_monitoring}
\end{figure*}

\section{Parameter Drift}
\label{sec:parameter_drift}

\subsection{Monitoring of control parameters}

During extended operation, the optimal values of control parameters drift through intrinsic device fluctuations, environmental changes, common reference variations, bias updates, and control-system drift~\cite{wellstood1987low, koch2007model, barends2013coherent}. Instead of recalibrating the full processor whenever performance changes, we quantify the calibration updates required by representative parameters and convert their relative stability into refresh priorities, without attributing each trajectory to a unique microscopic noise source.

The measurements were performed in a continuous region containing $26$ qubits and $36$ couplers. Within each cycle the qubit frequency $f_{01}$ is refreshed every hour from a Ramsey measurement. The $\pi$-pulse amplitude is then taken from a Rabi measurement, followed by the DRAG coefficient, the coupler idle bias, the CZ coupling $g$, and the CZ detuning; these parameters are refreshed every six hours. The cycle ends with the CZ dynamic phase, which is refreshed every hour. Each step runs only after the preceding parameter has been updated, so the recorded value of a later parameter depends on the change introduced by the previous step. In this schedule, $f_{01}$ and the dynamic phase were recorded over $44$ sampling iterations during a $44$-hour observation period, while the $\pi$-pulse amplitude, the DRAG coefficient, the coupler idle bias, the CZ coupling $g$, and the CZ detuning were recorded over a $48$-hour period with nine representative calibration iterations. The plotted trajectories therefore quantify the updates required at each measurement point rather than the noise spectrum of the underlying device.

The monitoring results are shown in Fig.~\ref{fig:parameter_monitoring}, where each trajectory is examined together with the response of its single-run calibration experiment to separate parameter variation from finite measurement resolution; for $f_{01}$, the shaded band indicates the $1\sigma$ repeatability of three Ramsey measurements per point. Several trajectories contain steps at the same iteration, consistent with common reference variations, bias updates, or dependencies of the sequential calibration procedure in addition to device fluctuations; the present records do not distinguish among these mechanisms. The measured update magnitudes relative to the repeatability and the calibration tolerance motivate the following classification.

\begin{table*}[htbp]
\centering
\caption{Drift metrics and refresh priority. }
\label{tab:drift_score}
\begin{tabular}{lcccc c}
\hline
Parameter & $\Delta$ & $\sigma$ & $D$  & $R$  & Priority \\
\hline
CZ coupling $g$ & $0.2005$~MHz & $0.07772$~MHz & 2.58 & 6.78\% & high \\
$f_{01}$ & $0.03140$~MHz & $0.007222$~MHz & 4.35 & 3.85\% & high \\
CZ dynamic phase & \SI{0.04354}{\rad} & \SI{0.03833}{\rad} & 1.14 & 2.80\% & high \\
DRAG coefficient & $0.03506$ & $0.01075$ & 3.26 & 0.492\% & medium \\
$\pi$-pulse amplitude & $0.00005279$ & $0.000038$ & 1.39 & 0.029\% & low \\
CZ detuning & $0.12376$~MHz & $0.25384$~MHz & 0.49 & 0.3\% & low \\
Coupler idle bias & $0.0001097$ & $0.001681$ & 0.065 & 0.002\% & low \\
\hline
\end{tabular}
\end{table*}

\subsection{Drift metrics and refresh rules}

To make the classification quantitative and transferable across devices, each record is referred to a common time $t_{\rm std}$. For a parameter value $x_{k,j}$ of device $k$ at elapsed time $t_{k,j}$ (in hours), the per-device drift magnitude is

\begin{equation}
\Delta_k=\sqrt{\frac{t_{\rm std}}{t_k}\sum_j (x_{k,j+1}-x_{k,j})^2},
\end{equation}
where $t_k$ is the total tracking time of device $k$ ($44$~h for $f_{01}$ and the dynamic phase, $48$~h for the other parameters) and $t_{\rm std}=1$~h is the reference time to which all drifts are scaled. 
 
To decide whether that drift is resolved, we define the detectability score

\begin{equation}
D=\frac{\Delta}{\sigma},
\end{equation}
where the per-point calibration uncertainty $\sigma$ is obtained by fitting the corresponding single-run calibration curve to the raw calibration data recorded during the same campaign. A value $D\ge1$ indicates that the drift exceeds the per-point uncertainty and is therefore resolvable.

To rank parameters with $D\ge 1$ by their effect on gate error, we use the sensitivity models of Appendix~\ref{app:sensitivity} to convert a parameter deviation into a gate error. The budget is $\epsilon_{\rm budget}=0.1\%$ for single-qubit gates and $1\%$ for two-qubit gates, and the budget fraction is

\begin{equation}
R=\frac{\epsilon_{\rm gate}(\Delta)}{\epsilon_{\rm budget}},
\end{equation}

where $R$ is dimensionless and $\epsilon_{\rm gate}(\Delta)$ is evaluated directly from those sensitivity models. Table~\ref{tab:drift_score} summarizes the measured metrics and the resulting ranking.

\subsection{Parameter hierarchy and stratified scheduling}

Applying these metrics to the monitored trajectories yields a three-tier hierarchy. The CZ effective coupling $g$, the qubit frequency $f_{01}$, and the CZ dynamic phase form the highest tier, with $R\approx6.78\%$, $3.85\%$, and $2.80\%$, respectively, and with $D=2.58$, $4.35$, and $1.14$. The CZ coupling $g$ has the largest budget contribution. The qubit frequency $f_{01}$ and the dynamic phase follow, consistent with the sensitivity of CZ phase accumulation to the qubit and coupler operating frequencies~\cite{sung2021realization}. These three parameters are assigned the highest refresh priority.

The DRAG coefficient forms the intermediate tier. Its one-hour-equivalent drift is clearly observable ($D=3.26$), but its budget contribution is moderate ($R\approx0.492\%$), so it is refreshed less frequently than the highest tier.

The $\pi$-pulse amplitude forms the low-priority active tier. Its one-hour-equivalent drift is observable ($D=1.39$), but its budget contribution is very small ($R\approx0.0290\%$), so it is refreshed only occasionally.

The CZ detuning and the coupler idle bias have $D<1$ ($D=0.49$ and $0.065$); their one-hour-equivalent drift is smaller than the per-point calibration uncertainty, so no truly observable drift exists. They are assigned to low priority and are re-examined periodically rather than scheduled for active refresh.

This ranking is established under the measured operating conditions and the available observation windows, and it supports differentiated refresh cadences without asserting a universal drift spectrum or long-term stability beyond the observations. The output of the analysis is a refresh priority ranking (high, medium, or low) rather than a numerical refresh time interval; the actual refresh cadence remains a separate scheduling choice informed by this priority. The same procedure can be generalized to other control parameters. The measurement uncertainty $\sigma$ can be determined from a single-run calibration curve or from the calibration history, and the one-hour-equivalent drift magnitude $\Delta$ is computed from the time-stamped calibration history. The ratio $D=\Delta/\sigma$ then determines whether the drift is observable, and $R=\epsilon_{\rm gate}(\Delta)/\epsilon_{\rm budget}$ sets the refresh priority.

We emphasize that the parameter set monitored here is illustrative rather than exhaustive. It covers the control parameters most relevant to the gate set used in this work, and the resulting hierarchy is specific to the present device, its architecture, and the measurement conditions under which the trajectories were recorded. Other processors, fabrication processes, or measurement environments may shift individual parameters between tiers or bring additional parameters into the resolved-drift regime. The contribution of this section is therefore not the particular ranking in Table~\ref{tab:drift_score} but the method itself, which can be applied to the calibration history of any processor. We offer it as a reference heuristic for the broader community.

\begin{figure*}[ht]
    \centering
    \includegraphics[width=1 \linewidth]{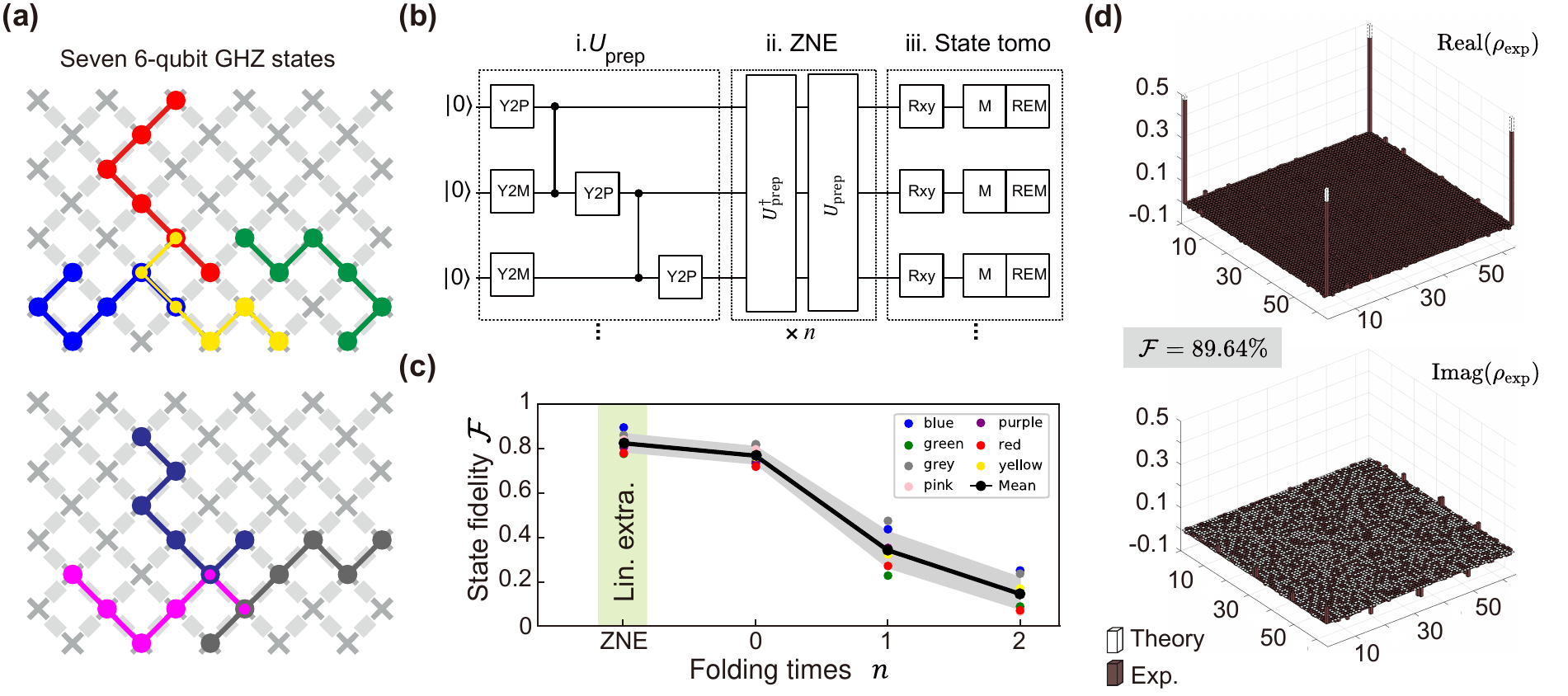}
    \caption{
Parallel preparation of the six-qubit GHZ state and the corresponding fidelity evaluation. (a) Seven independent one-dimensional $6$-qubit chains selected by unbiased sampling on the chip. (b) Experimental circuit combining zero-noise extrapolation (ZNE) and quantum state tomography. (c) Fidelity extrapolation curves for each test group under different folding times $n$. Different colored dots correspond to the different qubit chains in panel (a), and the shaded band denotes the spread across the seven chains. (d) Real and imaginary parts of the reconstructed density matrix for the best chain.
    }
    \label{fig:ghz_preparation}
\end{figure*}

\section{Validation}
\label{sec:validation}

To verify the effectiveness of the established parallel-density boundaries and parameter-refresh strategies at the system level, we deployed these strategies in a continuous testing region of the processor containing $26$ qubits and $36$ couplers, and performed GHZ state preparation and a single-cycle qubit-flip repetition code as benchmarks of performance in realistic computing scenarios. We further deployed the framework on a larger-scale superconducting quantum processor to test scalability. These benchmarks validate the calibrated control stack at the circuit level.

\subsection{GHZ State Preparation}
\label{subsec:ghz}
The GHZ state is a standard probe of coherence and multipartite entanglement~\cite{gisin1998bell,frowis2018macroscopic}. Its preparation is jointly sensitive to frequency drift, microwave-pulse errors, CZ conditional-phase and leakage errors, residual interactions, decoherence, and readout errors, and therefore tests at the circuit level whether the stratified refresh procedure maintains the mutually compatible conditions required for coherent multi-qubit evolution along a chain.

Seven one-dimensional $6$-qubit chains were sampled from the testing region [Fig.~\ref{fig:ghz_preparation}(a)], covering distinct local combinations of qubits, couplers, and readout resonators. Each chain was characterized by quantum state tomography combined with zero-noise extrapolation (ZNE) [Fig.~\ref{fig:ghz_preparation}(b)], with the circuit construction and analysis details given in Appendix~\ref{app:validation}. At the nominal circuit scale, the direct fidelities of the seven chains range from $71.92\%$ to $82.10\%$ and all exceed the $1/2$ fidelity witness threshold for genuine multipartite entanglement. Linear extrapolation of the measured curves [Fig.~\ref{fig:ghz_preparation}(c)] yields a largest state fidelity of $89.64\%$. The reconstructed density matrix of the best chain [Fig.~\ref{fig:ghz_preparation}(d)] retains both the endpoint populations and their mutual coherence without a systematic phase rotation (Appendix~\ref{app:validation}), which indicates that the refreshed qubit-frequency and CZ dynamic-phase parameters remain mutually compatible throughout the entangling sequence.

These results demonstrate that the stratified refresh strategy, applied after calibration at the parallel-density rules of Sec.~\ref{sec:parallel_calibration}, sustains complete six-qubit GHZ preparation across several local configurations of the calibrated region, rather than improving isolated gate metrics alone.

\subsection{Single-Cycle Qubit-Flip Repetition Code}
\label{subsec:repetition_code}

Quantum error correction circuits provide a demanding integrated test because ancilla preparation, entangling gates, and measurement must operate together. Error probabilities of decoded logical outcomes are widely used to characterize repetition-code experiments~\cite{wootton2018repetition,google2021exponential}. The present one-dimensional qubit-flip repetition code contains one stabilizer measurement layer followed by final readout. Here, it tests whether the stratified refresh strategy derived from parameter drift measurements sustains the calibrated conditions required for a parity-check circuit.
\begin{figure}[ht]
    \includegraphics[width=\linewidth]{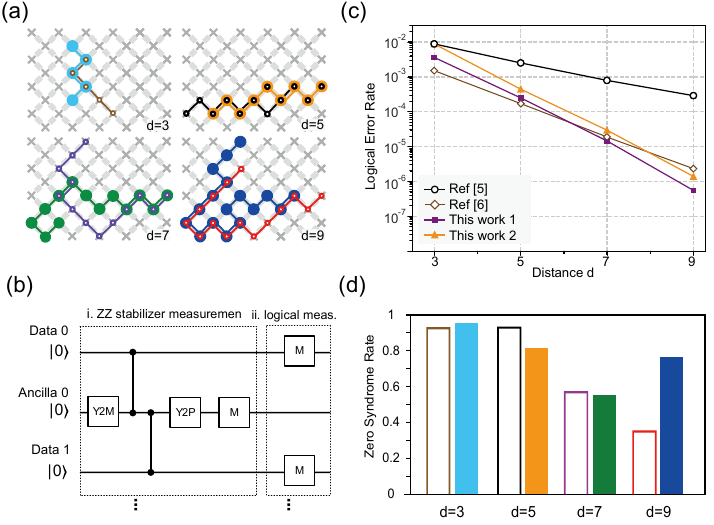}
    \caption{
    Implementation of the single-cycle qubit-flip repetition code and the scaling of the logical error rate. (a) Topology of the repetition code chains with distances $d=3,5,7,9$. Each distance group contains two different chains. (b) Parity-check circuit including state preparation, entanglement, and measurement. (c) Scaling of the logical error rate with code distance $d$. Purple squares and orange triangles are the two testing groups of this work. Open circles and diamonds are the surface-code logical error rates per cycle versus code distance reported by Refs.~\cite{google2023suppressing} and \cite{Google2024QEC}, respectively, and are shown only as an external reference; they are not the same metric as the single-cycle majority-vote error of this work. (d) Zero-syndrome rate for the different test groups at each code distance $d$.
    }
    \label{fig:qec_repetition_code}
\end{figure}

We selected two test group sequences at each code distance $d=3,5,7,9$ in the testing region [Fig.~\ref{fig:qec_repetition_code}(a)], with each circuit performing one ancilla-mediated nearest-neighbor parity check followed by a joint final measurement; the majority-vote decoding and the shot budget are defined in Appendix~\ref{app:validation}. The unconditional majority-vote error probabilities for the final $\lvert 0_L \rangle$ outcomes [Fig.~\ref{fig:qec_repetition_code}(c)] decrease by more than three orders of magnitude between $d=3$ and $d=9$ for both sequences. We note that this single-cycle majority-vote metric is dominated by state-preparation and final-readout errors rather than by the parity-check round, and is therefore not directly comparable to the logical error per cycle quoted in surface-code experiments~\cite{google2023suppressing,Google2024QEC}. Although the two sequences yield different absolute error probabilities, this common trend shows that the prioritized maintenance sustains the control conditions required for longer parity-check circuits while retaining the error suppression provided by redundant encoding.

The measured zero-syndrome rates, $P_{0}=\Pr(s_1=\cdots=s_{d-1}=0)$, are shown in Fig.~\ref{fig:qec_repetition_code}(d). A high $P_0$ is consistent with low aggregate error across the data qubits, ancilla qubits, and concurrent operations of a chain, and its variation across test chains provides an objective diagnostic of local hardware nonuniformity after the layered refresh procedure. Because the automated calibration framework traverses the full chip without performance-based selection, this variation is not hidden, and chains with depressed $P_0$ can be flagged for targeted recalibration or further hardware characterization. The refresh framework thereby doubles as a high-throughput diagnostic tool for the spatial performance of the processor.

\subsection{Calibration on a 337-Qubit Processor}
\label{subsec:504_qubit}

\begin{figure}[htbp]
    \centering
    \includegraphics[width=0.85\linewidth]{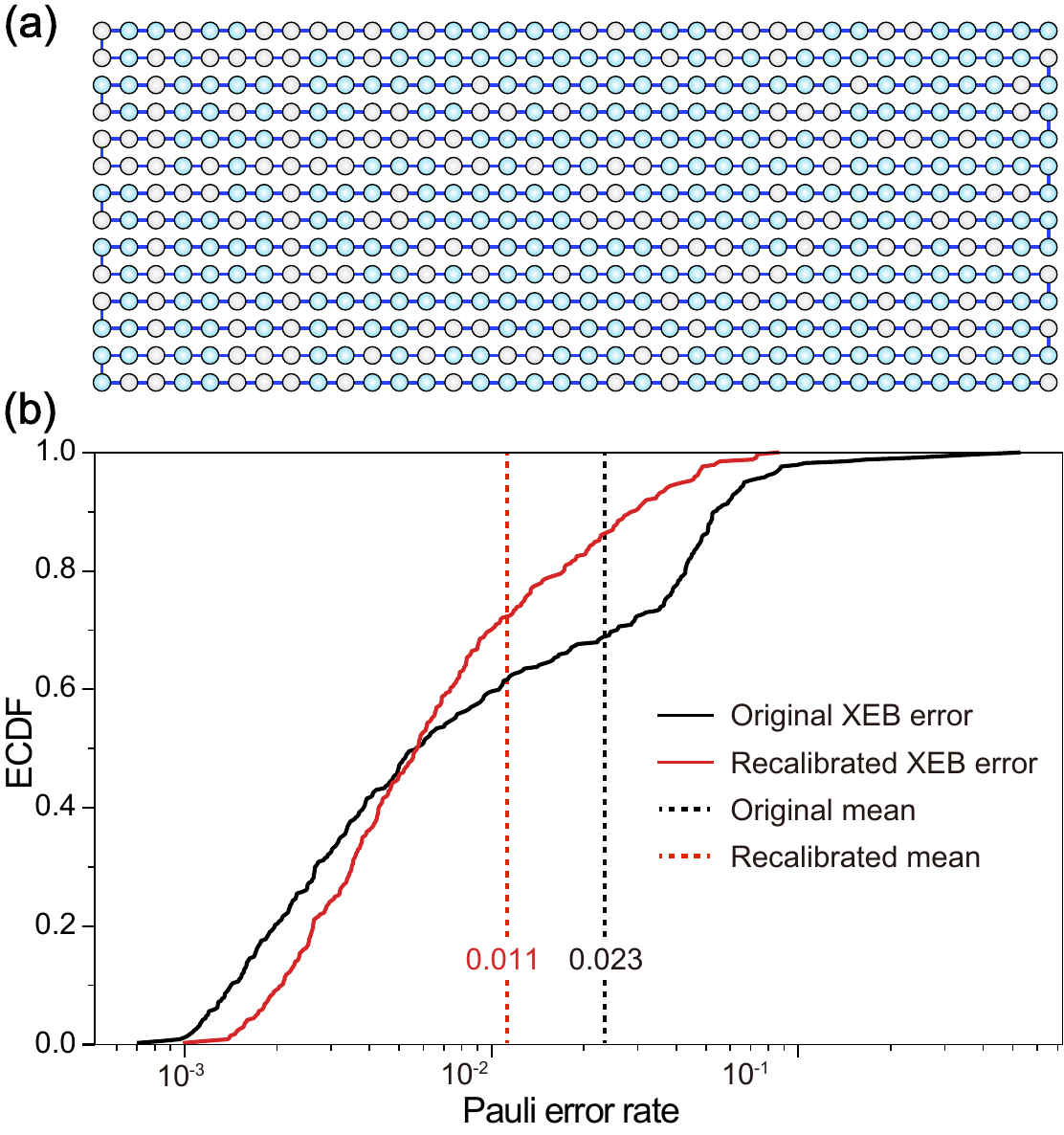}
    \caption{
    Scalability test on a large processor with 337 active qubits.
    (a) Topological layout of the processor, which contains 504 physical qubits (circles), with 337 active ones (blue circles). The connecting gray lines represent couplers. (b) ECDF of the single-qubit XEB Pauli error rates of the $337$ active qubits before (black) and after (red) recalibration.
    }
    \label{fig:scalability_504}
\end{figure}

To examine whether the workflow scales to larger processors, we deployed the automated calibration framework on a larger-scale processor (``Tianyan''), which uses fixed-frequency transmons with tunable couplers and operates $337$ active qubits [Fig.~\ref{fig:scalability_504}(a)].

The single-qubit XEB Pauli error rates obtained after the initial manual calibration and after automated recalibration are shown in Fig.~\ref{fig:scalability_504}(b). Averaged over the active qubits, the mean error decreases from $e_{\text{avg}}^{\text{pre}}=0.023$ to $e_{\text{avg}}^{\text{post}}=0.011$, corresponding to a $52\%$ reduction. The recalibrated distribution has a reduced high-error tail. The result therefore demonstrates an aggregate reduction in mean error and in the population of high-error qubits.

\section{Conclusion}
\label{sec:conclusion}

In this work, we have shown that the two central operational choices of automated calibration, namely how densely calibration tasks may be parallelized and when each control parameter should be refreshed, can be determined quantitatively from measurements on the processor itself rather than set by heuristic convention. Simultaneous single-qubit XEB leaves the Pauli error distribution nearly unchanged under full concurrency, while for CZ gates an exponential fit to the density-dependent control error gives a crosstalk decay length of $\lambda=1.37$ graph edges, consistently explained by flux crosstalk and effective-Hamiltonian analysis; these results establish $d_{\mathrm{Q}}=1$ and $d_{\mathrm{C}}=4$ as the densest tested configurations preserving XEB performance. Parameter monitoring further yields a refresh hierarchy quantified by the detectability score $D$ and the budget fraction $R$, separating the highest-priority parameters (CZ coupling $g$, qubit frequency $f_{01}$, and CZ dynamic phase) from the medium and low tiers, and scheduling each parameter by its measured drift and error-budget consumption rather than by universal wall-clock intervals. System-level experiments confirm the combined strategies, with six-qubit GHZ fidelities up to $89.64\%$ after zero-noise extrapolation, a repetition-code majority-vote error reduced by more than three orders of magnitude from $d=3$ to $d=9$, and the mean single-qubit error of $337$ qubits reduced from $0.023$ to $0.011$ by automated recalibration.

The demonstrated density rules and refresh strategy therefore serve as an experimentally grounded reference for the automated calibration of present-day processors and for the progressively autonomous maintenance required by future fault-tolerant operation. Future work will extend the observation period under continuous operation to test how the measured parameter hierarchy evolves over longer timescales, quantify the control, acquisition, and classical processing resources needed to preserve the parallel-calibration benefit as processors grow, and apply the same methodology to larger regions and longer circuit workloads.

\begin{acknowledgments}
We thank Ming Gong, Shaojun Guo, and Wenhao Chu for fruitful discussions. We are grateful to China Telecom Quantum Group for access to the ``Tianyan-504'' quantum computing cloud platform. This work was supported by the National Key R\&D Program of China (Grant No. 2024YFB4504000), the Fundamental and Interdisciplinary Disciplines Breakthrough Plan of the Ministry of Education of China (Grant No. JYB2025XDXM202), the Aid Program for Science and Technology Innovative Research Team in Higher Educational Institutions of Hunan Province, and the National Natural Science Foundation of China (Grant No. 62301572). XFZ acknowledges additional support from the National Science and Technology Major Project on Quantum Science and Technology (Grants No. 2025ZD0300600 and No. 2025ZD0300604).
\end{acknowledgments}

\vspace{5pt}
\textbf{Author contributions.}
MTD conceived the central concepts and coordinated the experimental work. ZHW, KCQZ, and SCX conducted the qubit experiments. ZHW, KCQZ, SCX, XFZ, ZDZ, and ZHY analyzed the experimental data. SH, DDL, PL, and ZHY developed the software under the supervision of XF. CL, DDL, and KDZ contributed to the experimental setup and the maintenance of the large-scale superconducting processor. MTD, JJW, and KL supervised both the theory and experiment. KCQZ, ZHW, SCX, and MTD co-wrote the manuscript with input from all authors.

\appendix

\section{Calibration Workflow Details}
\label{app:workflow}

The full-cycle workflow illustrated in Fig.~\ref{fig:qopilot_workflow} is representative of the calibration and maintenance procedure of large-scale superconducting processors. Starting from roughly estimated parameters at the initial cold start [Fig.~\ref{fig:qopilot_workflow}(a)], the processor is calibrated to a high-fidelity state [Fig.~\ref{fig:qopilot_workflow}(b)], and random parameter drift caused by environmental fluctuations subsequently displaces the operating point [Fig.~\ref{fig:qopilot_workflow}(c)], triggering maintenance updates. The directed acyclic graph (DAG) of the calibration experiments [Fig.~\ref{fig:qopilot_workflow}(d)] contains four core modules, namely, readout control, qubit control, coupler control, and environment characterization. In this work, the workflow is executed by the QoPilot automated calibration framework.

To achieve high-fidelity gates, we organized the calibration as the directed acyclic graph in Fig.~\ref{fig:qopilot_workflow}(d). We began with baseline readout, scanning resonator frequencies and calibrating the $Z$-spectrum to establish measurement capability. Once readout was operational, we moved to qubit control and ran coarse $f_{01}$ scans together with Rabi, Ramsey, and DRAG sequences to extract initial pulse parameters. The workflow then cycled back to readout pulse optimization and IQ classifier training, now using the preliminarily tuned qubits~\cite{wallraff2005approaching}. We iterated this loop until single-qubit $T_1$ and $T_2$ times~\cite{nakamura1999coherent, vion2002manipulating} and cross-entropy benchmarking fidelities stabilized~\cite{arute2019quantum, boixo2018characterizing}.

Next came single-qubit flux spectroscopy and idle bias calibration for the couplers, which we carried out via the QQSWAP operation. We then recalibrated the single-qubit gates to restore high fidelity. With those gates in place, we mapped the qubit frequency versus coupler bias voltage and fine-tuned the CZ core parameters, specifically the coupling strength $g$, the qubit detuning, and the dynamic phase~\cite{sung2021realization}. We assessed the overall calibration quality using two-qubit XEB gate fidelity.

\section{Analytical Model for Microwave Crosstalk in Single-Qubit Gates}
\label{app:microwave_crosstalk}

To quantify how parallel microwave driving affects single-qubit gates, we construct an analytical crosstalk model. Under the rotating-wave approximation (RWA), the Hamiltonian of a target qubit in the rotating frame, incorporating the spatial microwave leakage from a concurrently driven neighboring qubit, is given by~\cite{wang2022control,plenio1998quantum}
\begin{equation}
\begin{split}
H_{d}(t)&= \frac{1}{2} \Omega(t) \sigma_x + \frac{1}{2}\eta \Omega(t) \big[ \cos(\Delta t) \sigma_x + \sin(\Delta t) \sigma_y \big] \\
&\quad - i \begin{pmatrix} \frac{1}{T_\phi} & 0 \\ 0 & \frac{1}{2T_1} \end{pmatrix},
\end{split}
\end{equation}
where $\sigma_x$ and $\sigma_y$ are the standard Pauli matrices. The intended driving field is parameterized by a raised-cosine envelope $\Omega(t) = \frac{A}{2}\left[1 - \cos\left(\frac{2\pi t}{\tau}\right)\right]$, with $A$ being the peak driving amplitude and $\tau$ the gate duration. The microwave crosstalk is characterized by a dimensionless coefficient $\eta$ and a frequency detuning $\Delta$ between the target and the adjacent victim qubits. The non-Hermitian term describes the environment-induced decoherence, parameterized by the energy relaxation time $T_1$ and the pure dephasing time $T_\phi$.

By integrating the time-dependent Hamiltonian over the gate duration $\tau$, the actual evolution matrix $U_{\text{actual}} = e^{-i \int_0^\tau H_{d}(t) dt}$, which inherently captures both the driving crosstalk and decoherence, can be analytically expressed as
\begin{equation}
U_{\text{actual}} = e^{-\frac{\tau}{2T_2}} \begin{pmatrix} \cos B - \frac{\gamma_- \tau}{B}\sin B & -i\frac{(\frac{\pi}{2} + \delta_x) - i \delta_y}{B}\sin B \\ -i\frac{(\frac{\pi}{2} + \delta_x) + i \delta_y}{B}\sin B & \cos B + \frac{\gamma_- \tau}{B}\sin B \end{pmatrix}.
\end{equation}
Here, $T_2 = (\frac{1}{2T_1} + \frac{1}{T_\phi})^{-1}$ is the coherence time, $\gamma_- = \frac{1}{2}(\frac{1}{T_\phi} - \frac{1}{2T_1})$, and the generalized complex angle is defined as $B = \sqrt{ (\frac{\pi}{2} + \delta_x)^2 + \delta_y^2 - \gamma_-^2 \tau^2 }$. The microwave leakage introduces parasitic driving errors during the gate operation, which can be explicitly quantified by the time integrals $\delta_x = \int_0^\tau \frac{1}{2}\eta \Omega(t) \cos(\Delta t) dt$ and $\delta_y = \int_0^\tau \frac{1}{2}\eta \Omega(t) \sin(\Delta t) dt$.

The average gate fidelity is analytically evaluated by comparing the actual evolution against the ideal target matrix $U_{\text{ideal}}$~\cite{nielsen2002simple}
\begin{equation}
F = \frac{1}{d^2} \left| \text{Tr}\left(U_{\text{ideal}}^\dagger U_{\text{actual}}\right) \right|^2,
\end{equation}

where $d=2$ is the dimension of the single-qubit Hilbert space. Within this model, the fidelity penalty is governed by the crosstalk-induced amplitude error $\delta_x$ and the rotation-axis tilt $\delta_y$. As the spatial distance between driven qubits increases, the crosstalk strength $\eta$ attenuates rapidly and both $\delta_x$ and $\delta_y$ approach zero. In this limit, crosstalk-induced interference is fully suppressed, and the operational fidelity is bounded only by the intrinsic coherence limit.

\section{Effective Hamiltonian and Fidelity Analysis for Flux Crosstalk in Two-Qubit Gates}
\label{app:flux_crosstalk}

To quantify the degradation of two-qubit CZ gates under parasitic flux crosstalk, we analyze the induced interference with an effective Hamiltonian model. For a pair of target qubits coupled via an intermediate tunable coupler, the coupler mode can be adiabatically eliminated using a Schrieffer-Wolff transformation. In the dispersive regime, the resulting effective two-qubit Hamiltonian is given by~\cite{zhao2020high}
\begin{equation}
\begin{split}
H_{\text{eff}} &= \tilde{\omega}_1 |10\rangle\langle 10| + \tilde{\omega}_2 |01\rangle\langle 01| \\
&\quad + \big(\tilde{\omega}_1 + \tilde{\omega}_2 + \zeta(\omega_c)\big) |11\rangle\langle 11| \\
&\quad + g_{\text{eff}}(\omega_c) (\sigma_1^+ \sigma_2^- + \text{h.c.}),
\end{split}
\end{equation}
where $\tilde{\omega}_1$ and $\tilde{\omega}_2$ are the dressed frequencies of the two qubits, $|mn\rangle$ denotes the joint computational basis, and $\sigma_j^\pm$ represents the raising and lowering operators for the $j$-th qubit. During a CZ gate operation, the conditional phase accumulation rate (i.e., the $ZZ$ interaction strength) $\zeta(\omega_c)$ is dictated by the effective transverse coupling strength $g_{\text{eff}}(\omega_c)$,
\begin{equation}
\begin{split}
g_{\text{eff}}(\omega_c) &\approx g_{12} + \frac{g_{1c} g_{2c}}{2} \left( \frac{1}{\omega_1 - \omega_c} + \frac{1}{\omega_2 - \omega_c} \right), \\
\zeta(\omega_c) &\approx \frac{2 g_{\text{eff}}(\omega_c)^2}{\Delta_{12} - \alpha_2}.
\end{split}
\end{equation}
Here, $g_{12}$ is the direct capacitive coupling strength between the two qubits, whereas $g_{1c}$ and $g_{2c}$ denote the coupling strengths between each qubit and the tunable coupler. The parameters $\omega_1$, $\omega_2$, and $\omega_c$ are the bare frequencies of the corresponding components at the operating point, $\Delta_{12} \equiv \omega_1 - \omega_2$ is the qubit-qubit frequency detuning, and $\alpha_2$ is the anharmonicity of the qubit excited to the $|2\rangle$ state.

Because the tunable coupler is biased away from its flux sweet spot to mediate the CZ interaction, its operating frequency $\omega_c$ becomes highly sensitive to external magnetic flux variations~\cite{fu2026flux}. Spatial cross-leakage of flux-control signals, arising from parasitic mutual inductance between adjacent control lines, manifests as a continuous low-frequency drift $\delta \omega_c$ in the coupler's base frequency. This drift modifies the effective coupling and shifts the conditional-phase rate by $\delta \zeta \approx \frac{\partial \zeta}{\partial \omega_c} \delta \omega_c$. Integrating this deviation over the gate duration $\tau$, the actual evolution matrix $U_{\text{actual}} = e^{-i \int_0^\tau H_{\text{eff}}(t) dt}$ accumulates a spurious conditional phase error $\delta \Theta$,
\begin{equation}
\delta \Theta = \int_0^\tau \delta \zeta(t) dt \approx \left( \frac{\partial \zeta}{\partial \omega_c} \right) \cdot \delta \omega_c \cdot \tau.
\end{equation}

For a two-qubit Hilbert space ($d=4$), the average gate fidelity penalized by this unintended conditional phase error can be analytically evaluated against the ideal CZ matrix $U_{\text{ideal}} = \text{diag}(1, 1, 1, -1)$,
\begin{equation}
F = \frac{1}{d^2} \left| \text{Tr}\left(U_{\text{ideal}}^\dagger U_{\text{actual}}\right) \right|^2 \approx 1 - \frac{3}{20} (\delta \Theta)^2.
\end{equation}
This derivation reveals a quadratic fidelity penalty $1-F \propto (\delta\omega_c)^2$ with respect to the flux-induced frequency drift. Consequently, ensuring a sufficient spatial separation to physically attenuate the mutual inductance and minimize $\delta\omega_c$ is essential to bound the conditional phase error $\delta\Theta$ and preserve high fidelity in parallel multi-qubit architectures.

\section{Sensitivity Models for Calibration-Parameter Tolerances}
\label{app:sensitivity}

The refresh-priority score in Sec.~\ref{sec:parameter_drift} requires the fidelity tolerance $\tau_i$ of each control parameter, i.e., the parameter deviation that consumes a prescribed gate-error budget $\epsilon_{\rm budget}$. We set $\epsilon_{\rm budget}=10^{-3}$ for single-qubit gates and $\epsilon_{\rm budget}=10^{-2}$ for two-qubit gates, consistent with the measured Pauli-error scales of this device. For a parameter with a first-order sensitivity $s_i=|\partial \epsilon_{\rm gate}/\partial p_i|$, the tolerance is
\begin{equation}
\tau_i=\frac{\epsilon_{\rm budget}}{s_i}.
\end{equation}
For a parameter whose working point is a minimum of the error function, the leading dependence of the error is quadratic,
\begin{equation}
\epsilon_{\rm gate}(p_i)\approx \epsilon_0+\kappa_i (p_i-p_i^{\rm opt})^2,
\qquad
\tau_i=\sqrt{\frac{\epsilon_{\rm budget}}{\kappa_i}}.
\end{equation}
The measured calibration scans provide $s_i$ or $\kappa_i$ directly at the operating point.

\textbf{$\pi$-pulse amplitude.} The single-qubit excitation probability follows $P_1=\sin^2(\pi A/2A_\pi)$. For a small amplitude deviation $\delta A$,
\begin{equation}
\epsilon\approx \left(\frac{\pi\,\delta A}{2A_\pi}\right)^2,
\qquad
\tau_A=\frac{2A_\pi}{\pi}\sqrt{\epsilon_{\rm budget}}.
\end{equation}
With the measured median $A_\pi=0.153$ and $\epsilon_{\rm budget}=10^{-3}$, $\tau_A=3.1\times10^{-3}$.

\textbf{Qubit frequency $f_{01}$.} For a resonant single-qubit gate with Rabi frequency $\Omega/2\pi\approx5$~MHz, a detuning $\delta f$ produces a gate error $\epsilon\approx(2\pi\,\delta f/\Omega)^2$, giving
\begin{equation}
\tau_f=\frac{\Omega}{2\pi}\sqrt{\epsilon_{\rm budget}}.
\end{equation}
With $\epsilon_{\rm budget}=10^{-3}$, $\tau_f\approx0.16$~MHz. The same frequency detuning also enters the CZ conditional phase, so $f_{01}$ is treated as the common upstream source of the dynamic-phase excursions.

\textbf{CZ dynamic phase.} From the phase-error fidelity relation in Appendix~\ref{app:flux_crosstalk},
\begin{equation}
\epsilon\approx\frac{3}{20}(\delta\Theta)^2,
\qquad
\tau_\Theta=\sqrt{\frac{20\epsilon_{\rm budget}}{3}}.
\end{equation}
With $\epsilon_{\rm budget}=10^{-2}$, $\tau_\Theta=\SI{0.26}{\rad}$.

\textbf{CZ effective coupling $g$.} The leakage measured in the $g$-calibration scan has a parabolic minimum at $g^{\rm opt}$. A quadratic fit near the operating point gives $\kappa_g=0.017$~MHz$^{-2}$, so
\begin{equation}
\tau_g=\sqrt{\frac{\epsilon_{\rm budget}}{\kappa_g}}.
\end{equation}
With $\epsilon_{\rm budget}=10^{-2}$, $\tau_g=0.77$~MHz.

\textbf{CZ detuning.} The conditional-phase error measured in the detuning scan is approximately linear near the operating point with slope $s_d=\SI{0.114}{\rad\per\mega\hertz}$. Combining the linear sensitivity with the quadratic phase-error-to-infidelity relation gives
\begin{equation}
\tau_d=\frac{1}{s_d}\sqrt{\frac{20\epsilon_{\rm budget}}{3}}.
\end{equation}
With $\epsilon_{\rm budget}=10^{-2}$, $\tau_d=2.26$~MHz.

\textbf{Coupler idle bias.} The qubit-qubit swap population measured in the QQSWAP scan is parabolic near the idle point. The observed idle-bias excursions are smaller than the calibration step size, so the associated risk is negligible.

\textbf{DRAG coefficient.} The DRAG calibration-response fits give an estimated repeatability of $\sigma_{\rm DRAG}\approx 0.011$. The one-hour-equivalent drift of the DRAG coefficient is observable ($D=3.26$) but its budget contribution is moderate, so the risk remains in the medium-priority class.

The resulting tolerances $\tau_i$ are listed in the parameter paragraphs above. For the one-hour-equivalent drift magnitude $\Delta_i$ defined in Sec.~\ref{sec:parameter_drift}, the budget fraction consumed by the drift is
\begin{equation}
R_i=\frac{\epsilon_{\rm gate}(\Delta_i)}{\epsilon_{\rm budget}},
\end{equation}
with $R_i=\Delta_i/\tau_i$ for first-order sensitivities and $R_i=(\Delta_i/\tau_i)^2$ for the quadratic models above. The detectability gate $D_i=\Delta_i/\sigma_i$ defined in Sec.~\ref{sec:parameter_drift} compares the one-hour-equivalent drift with the per-point calibration uncertainty $\sigma_i$. Parameters with $D_i<1$ are assigned to low priority; the remaining parameters are ranked by decreasing $R_i$, giving the high, medium, and low priority ordering in Table~\ref{tab:drift_score}.

\section{Details of the Validation Benchmarks}
\label{app:validation}

\subsection{GHZ tomography and zero-noise extrapolation}

The ZNE and tomography procedure used in Sec.~\ref{subsec:ghz} is illustrated in Fig.~\ref{fig:ghz_preparation}(b). Let $U_{\mathrm{prep}}$ denote the GHZ preparation circuit. For the integer folding number $n$, the experiment implements $U_{\mathrm{prep}}(U_{\mathrm{prep}}^{\dagger}U_{\mathrm{prep}})^n$, followed by local $R_{xy}$ rotations, measurement, and readout error mitigation. This global unitary-folding construction preserves the ideal target state while increasing the physical depth from one to $2n+1$ GHZ circuit segments, and is a standard realization of ZNE. We use linear extrapolation throughout. Folded versions of the preparation circuit provide the different scaling factors, and for each scaling factor we perform full quantum state tomography on the output state, collecting measurement outcomes in the required bases. Linear extrapolation of each measured quantity then gives the mitigated zero-noise estimate
$p_{i}^{\mathrm{EM}}=\mathrm{ZNE}\left(p_{i}^{(s_1)},p_{i}^{(s_2)},\ldots,p_{i}^{(s_k)},\ldots\right)$,
where $p_{i}^{(s_k)}$ denotes the estimate of observable $i$ at scaling factor $s_k$. The extrapolated observables together define the ZNE-mitigated density matrix~\cite{temme2017error,he2020zero,zhang2020error}. The real part of the reconstructed $\rho_{\mathrm{exp}}$ for the chain yielding the largest extrapolated value [Fig.~\ref{fig:ghz_preparation}(d)] contains the two dominant GHZ populations at $\lvert 0^6\rangle$ and $\lvert 1^6\rangle$ together with the corresponding corner coherence terms, while the imaginary panel contains no comparably large structured component. Writing $P_6=\rho_{0^6,0^6}+\rho_{1^6,1^6}$ and $C_6=2\operatorname{Re}(\rho_{0^6,1^6})$ for the target phase used here gives $F_{\mathrm{GHZ}}=(P_6+C_6)/2$, which separates the population retained in the target GHZ subspace from the coherence of the superposition. As the number of folds increases, $C_6$ decreases more rapidly than $P_6$, while the phase of the corner coherence remains close to its initial value instead of accumulating a monotonic rotation. The nonzero corner coherence and the absence of a systematic phase rotation show that the nominal-depth state retains a coherent superposition in the target GHZ subspace, rather than only the endpoint populations.

\subsection{Repetition-code decoding}

Each circuit in Sec.~\ref{subsec:repetition_code} yields $d$ data outcomes and $d-1$ syndrome outcomes per shot, and the final data outcomes are decoded by a majority vote. Writing the measured data bits as $b_i\in\{0,1\}$, we define the $\lvert 0_L \rangle$ majority-vote error probability as
\begin{equation}
\epsilon_{L,0}=\Pr\left(\sum_{i=1}^{d} b_i > \frac{d-1}{2}\right).
\end{equation}
No zero-syndrome postselection is applied, and the syndrome record is retained as diagnostic information only. The number of shots per point increases from $10^{6}$ at $d=3$ to $10^{8}$ at $d=9$ so that rare majority-vote errors remain resolved at the largest distances.

\bibliography{references_revised}

\end{document}